\documentclass[superscriptaddress,amsmath,amssymb, aps, prx,longbibliography,twocolumn, footinbib]{revtex4-2}

\usepackage{upgreek}
\usepackage{color}
\usepackage{graphicx}
\usepackage{nccmath}
\usepackage{bm}
\usepackage{hyperref}
\hypersetup{colorlinks,breaklinks,
            urlcolor=[rgb]{0,0,0.36},
            linkcolor=[rgb]{0,0,0.36},
            citecolor=[rgb]{0,0,0.36}}
\usepackage[mathlines]{lineno}
\usepackage{dcolumn}
\usepackage{mathtools}  
\usepackage{siunitx}    
\usepackage{dsfont}

\usepackage{epsfig,tabularx,multirow,booktabs}

\usepackage{braket}

\usepackage[normalem]{ulem}
\newcommand{\comment}[1]{}			

\definecolor{cadmiumgreen}{rgb}{0.0, 0.42, 0.24} 

\usepackage{pifont} 
\newcommand{\xmark}{\ding{55}}
\newcommand{\cmark}{\ding{52}}

\usepackage{adjustbox} 
\usepackage{tabularx} 
\newcolumntype{Y}{>{\centering\arraybackslash}X} 

\newcounter{SN}

\usepackage{array}
\newcolumntype{M}[1]{>{\centering\arraybackslash}m{#1}}
\makeatletter
\renewcommand\footnotesize{%
   \@setfontsize\footnotesize\@ixpt{8}%
   \abovedisplayskip 8\p@ \@plus2\p@ \@minus4\p@
   \abovedisplayshortskip \z@ \@plus\p@
   \belowdisplayshortskip 4\p@ \@plus2\p@ \@minus2\p@
   \def\@listi{\leftmargin\leftmargini
               \topsep 4\p@ \@plus2\p@ \@minus2\p@
               \parsep 2\p@ \@plus\p@ \@minus\p@
               \itemsep \parsep}%
   \belowdisplayskip \abovedisplayskip
}
\makeatother

\newcommand{\Harvard}{Department of Physics, Harvard University, Cambridge, Massachusetts 02138, USA.}

\newcommand{\HarvardAppl}{John A. Paulson School of Engineering and Applied Sciences, Harvard University, Cambridge, MA 02138.}

\newcommand{\MaxP}{Max Planck Institute for the Structure and Dynamics of Matter, Luruper Chaussee 149, 22761 Hamburg, Germany.}

\newcommand{\Oxford}{Clarendon Laboratory, University of Oxford, Parks Road, Oxford OX1 3PU, UK.}

\newcommand{\ETH}{Institute for Theoretical Physics, ETH Zurich, 8093 Zurich, Switzerland.}

\begin{document}


\title{Optically-induced Umklapp shift currents in striped cuprates}
\author{Pavel~E.~Dolgirev}
\email{p\_dolgirev@g.harvard.edu}
\affiliation{\Harvard}

\author{Marios~H.~Michael}
\affiliation{\Harvard}

\author{Jonathan~B.~Curtis}
\affiliation{\Harvard}
\affiliation{\HarvardAppl}

\author{Daniel~E.~Parker}
\affiliation{\Harvard}

\author{Daniele~Nicoletti}
\affiliation{\MaxP}
\author{Michele~Buzzi}
\affiliation{\MaxP}
\author{Michael~Fechner}
\affiliation{\MaxP}

\author{Andrea~Cavalleri}
\affiliation{\MaxP}
\affiliation{\Oxford}
\author{Eugene~Demler}
\affiliation{\ETH}

\date{\today}

\begin{abstract}
Motivated by recent experiments that observed low-frequency second-order optical responses in doped striped superconductors, here we investigate the nonlinear electrodynamics of systems exhibiting a charge density wave (CDW) order parameter. Due to the Bragg scattering off the CDW order, an incoming spatially homogeneous electric field in addition to zero momentum current generates Umklapp currents that are modulated in space at momenta of the reciprocal CDW lattice. In particular, here we predict and microscopically evaluate the Umklapp shift current, a finite momentum analog of the regular shift current which represents the second-order optical process that downconverts homogeneous AC electric field into low-frequency, zero momentum current. Specifically, we evaluate real-time response functions within mean-field theory via the Keldysh technique and use the Peierls substitution to compute observables at finite momenta in lattice models. We find that systems with certain lattice symmetries (such as inversion symmetry), where the regular shift current is disallowed, may give rise to the Umklapp one. We apply our framework to investigate lattice symmetries in layered materials with helical-like stripes and show that both types of shift currents provide insight into the nature of intertwined phases of matter. Finally, we discuss the relation of our findings to recent experiments in striped superconductors.
\end{abstract}

\maketitle

\section{Introduction}

Strongly correlated materials, such as high-$T_c$ cuprate superconductors, typically feature a rich phase diagram, exhibiting a complex interplay of superconducting, spin, charge, and structural orders~\cite{RevModPhys.87.457}. A canonical example is the family of La$_{2-x}$Ba$_x$CuO$_4$ cuprates that exhibit both stripe charge and superconducting orders~\cite{Berg_2009,Berg.2009.b,Berg.2009.c}. Interestingly, in these systems, the stripes have no clear signatures in the linear optical measurements~\cite{Rajasekaran.2018}. However, the interplay of the two orders can lead to unique nonlinear electromagnetic responses, and, in particular, the emergence of a giant third harmonic in the out-of-plane nonlinear terahertz reflectivity was suggested as evidence for superfluid stripes~\cite{Rajasekaran.2018}. Motivated at exploring peculiar responses in the electrodynamics of cuprates, 
there has been a renewed effort in developing and applying nonlinear and pump-probe spectroscopies, and recent experimental progress has lead to the detection of inversion symmetry breaking via second-harmonic generation~\cite{Zhao.2017,deLaTorre.2021}, observation of charge density wave sliding motion~\cite{Mitrano.2019}, and detection of terahertz emission in photoexcited striped superconductors~\cite{stripes1}.

This latter experiment in the doped striped cuprate superconductor La$_{2-x}$Ba$_x$CuO$_4$ poses several puzzles that require careful theory examinations~\cite{stripes1}. The experimental findings can be summarized as follows. Upon strong photoexcitation at optical frequency, coherent outgoing radiation was observed, with the frequency being sharply peaked at the Josephson plasma resonance, which is at least two orders in magnitude smaller than the pump pulse frequency. Notably, this effect is present only in the phase where both superconducting and stripe orders coexist. It was argued in Refs.~\cite{stripes1,stripes2} that behind the experiment is a second-order optical nonlinearity in the form of shift current, which is activated due to the stripe order. More precisely, under the assumption that stripes give rise to a shift current, it was demonstrated in Ref.~\cite{stripes2} that the interplay between this current and low-momenta surface Josephson plasmons results in sharp in frequency radiation, consistent with the experiment. However, while stripes in cuprates are known to exhibit various linear and nonlinear optical effects~\cite{Homes.2006,Wen.2012,PhysRevB.78.174529,Li.2007,Moore.2013,Golub.2020,Hosur2013,Dienst.2013,Cremin.2021}, in accordance with symmetry considerations, one does not expect the emergence of shift currents, at least for the simplest models of stripes~\cite{Tranquada.2012,Tranquada.2020,Comin.2015,Comin.2016}.  Motivated to resolve this issue, we develop a general theory for the nonlinear electromagnetic response of materials with a charge density wave (CDW) order parameter. We subsequently apply this formalism to investigate layered systems with stripes.

A special feature of materials with a CDW order, which will be crucial to our subsequent discussion, is that zero momentum pump pulse can generate not only homogeneous currents but also the Umklapp ones at momenta of the CDW lattice, resulting in unique signatures in nonlinear responses. Therefore nonlinear processes in CDW systems are characterized not only by frequency mixing~\cite{butcher1991elements,Boyd,Bloembergen,ma2021topology,ahn2021riemannian,Parker2019Diagrammatic,morimoto2016topological,dolgirev2021periodic,Kaneko.2021,Orenstein.2021} but also by Bragg momenta mixing. Remarkably, as the semi-phenomenological analysis of Ref.~\cite{stripes2} indicated, regular homogeneous shift current is insufficient to explain the terahertz emission experiment~\cite{stripes1}, and an appreciable Umklapp shift current is required. Motivated to provide a microscopic foundation behind the discussions of Refs.~\cite{stripes1,stripes2}, here we develop a general microscopic theory for such low-frequency Umklapp shift current, the central focus of the paper. By investigating various charge order patterns of striped materials, we argue that the Umklapp shift current brings insight into the understanding of intertwined phases of matter. We note that this current can be experimentally detected (we briefly discuss this in Sec.~\ref{subsec:exp}), even with far-field measurements.

The paper is structured as follows: In Sec.~\ref{sec:general}, we formulate a general theory for perturbative evaluation of electrodynamic properties in CDW materials, where the effects of the electronic bandstructure back folding play a crucial role. Then, in sections~\ref{sec:linear} and~\ref{sec:nonlinear}, we evaluate linear and second-order electromagnetic responses, respectively, and, in particular, provide the expression~\eqref{eqn:sigma_shiftU} for the Umklapp shift current. In Sec.~\ref{sec:striped}, we apply the developed framework to investigate various lattice symmetries and their implications for the second-order response in layered materials with helical-like stripes. Intriguingly, we find certain patterns of stripes cannot give rise to the regular shift current, but can host the Umklapp one. We point out that the putative shift currents observed in the recent photoexcitation experiments in striped superconductors~\cite{stripes1} provide insights into the structure of CDW and superconducting orders in these systems~\cite{Lee.2021}. Finally, we briefly summarize our findings and discuss the outlook for future directions in Sec.~\ref{sec:conclusion}.

\section{Theoretical framework}
\label{sec:general}

In this section, we set up the formalism for evaluating the nonlinear electromagnetic response of systems exhibiting a CDW order parameter. We remark that our derivation closely follows the diagrammatic approach of Ref.~\cite{Parker2019Diagrammatic}. We differ essentially in two aspects. The first one is that we employ the Keldysh technique to evaluate various real-time response functions; this generalization, in addition to its convenience, provides a more transparent and physical structure of the perturbative expansion. More essential is that we extend the previous work to evaluate the responses at a finite momentum, such as the Umklapp shift current, which we discuss in the following sections.

\subsection{Operators in the presence of the external homogeneous electric field}

Here we treat the effects of external homogeneous (zero momentum) pump pulses perturbatively and work in the so-called velocity gauge~\cite{von1981theory,passos2018nonlinear,Parker2019Diagrammatic}. In systems with a CDW order, such a pump pulse generally generates electric fields both at zero momentum and at momenta of the CDW reciprocal lattice. Below we are interested in evaluating the response of an electric current at a finite momentum to the zero momentum electric field only. To evaluate any observables, we first need to provide the corresponding expressions for these observables in the presence of the external homogeneous applied field. To this end, we consider the following microscopic model, where the external electromagnetic field is incorporated via the Peierls substitution (throughout the paper, we set $\hbar = c = k_B = 1$):
\begin{align}
    \hat{H} = \hat{H}_{\rm kin} + {\hat H}_{\rm int},
\end{align}
where the kinetic energy is given by
\begin{align}
    \hat{H}_{\rm kin} = -\sum_{ij} t_{ij} \hat{c}^\dagger_i\hat{c}_j \exp\Big( -i e \int_{\bm r_i}^{\bm r_j} d\bm r \cdot  \bm A(\bm r,t)  \Big).
\end{align}
Here $t_{ij}$ encode hopping elements on a cubic lattice, $e$ is the electric charge, and $\bm A(\bm r,t)$ is the vector potential. $\hat{c}^\dagger_i$ and $\hat{c}_i$ are the fermionic creation and annihilation operators at site $i$, respectively. At this stage, we do not need to specify the interaction Hamiltonian $\hat{H}_{\rm int}$; however, for the presented framework to be consistent, one requires that the Peierls substitution does not affect $H_{\rm int}$ (an example of an allowed Hamiltonian is the density-density Coulomb interaction).

The electric current density can be evaluated from the derivative of the kinetic Hamiltonian $\hat{H}_{\rm kin}$ with respect to the vector potential in the usual manner. Specifically, to get the current density at a finite momentum $\bm Q = Q\hat{z}$, we write the vector potential as: $\bm A(\bm r, t)  = \bm A_0(t) + \delta \bm A_Q(\bm r, t)$, where $\bm A_0(t)$ is the homogeneous component and $\delta\bm A_Q(\bm r, t)$ represents the harmonic at momentum $\bm Q$:
\begin{align}
    \delta\bm  A_Q(\bm r,t) = \Big(A_Q(t)e^{i Q z} + A^*_Q(t)e^{-i Q z} \Big)\hat{z}.
\end{align}
Then the current operator at $\bm Q$ and pointing along the $z$-axis is given by:
\begin{align}
    \hat{J}_Q & \equiv -\frac{\delta \hat{H}_{\rm kin} }{\delta A_Q^*}\Bigg|_{ A_Q = 0} \notag\\
    &= \frac{e}{Q} \sum_{ij} t_{ij} \hat{c}^\dagger_i\hat{c}_j \Big[ e^{-i \bm Q\cdot \bm r_j} - e^{-i \bm Q\cdot \bm r_i} \Big]\notag\\
    &\qquad\qquad\qquad\quad
    \times \exp\Big( -i e (\bm r_j - \bm r_i)\cdot \bm A_0(t) \Big) . \label{eqn:current_Q}
\end{align}
Importantly, Eq.~\eqref{eqn:current_Q} includes the homogeneous component $\bm A_0(\bm r,t)$ to arbitrary order, which in turn will allow us to evaluate the corresponding nonlinear responses accurately. For concreteness, below we assume $\bm A_0(t) = A_0(t) \hat{z}$. In this case, while the zero momentum current operator can be obtained similarly from evaluating the variational derivative with respect to $A_0$, it can also be obtained from Eq.~\eqref{eqn:current_Q} by taking the limit $Q\to 0$. The presented framework can be straightforwardly generalized to arbitrary directions of both the external field and electric current.

Here we primarily investigate the electrodynamic properties of layered materials. For this reason and simplicity, we assume that the coupling between layers takes the form of nearest-neighbor hopping $t_z$ (the discussion below can be correspondingly modified if this coupling between the layers is different). Aiming at evaluating the second-order response, we now expand the kinetic Hamiltonian and the current operators up to the second-order in $A_0(t)$ and write them in momentum space:
\begin{widetext}
\begin{gather}
    \hat{H}_{\rm kin}  = \int \frac{d^3 \bm k}{(2\pi)^3} H_0 \hat{c}^\dagger_{\bm k }\hat{c}_{\bm k} =  \int \frac{d^3 \bm k}{(2\pi)^3} \epsilon(\bm k) \hat{c}^\dagger_{\bm k }\hat{c}_{\bm k}  - 2 e t_z\int \frac{d^3 \bm k}{(2\pi)^3} \Big[ A_0(t) \sin k_z -\frac{1}{2} eA^2_0(t) \cos k_z
    \Big] \hat{c}^\dagger_{\bm k }\hat{c}_{\bm k}  + \dots,
    \\
    {\hat J}_0  = 2 e t_z\int \frac{d^3 \bm k}{(2\pi)^3} \Big[ \sin k_z - eA_0(t) \cos k_z
    - \frac{1}{2}e^2 A^2_0(t)\sin k_z \Big] \hat{c}^\dagger_{\bm k }\hat{c}_{\bm k} + \dots,\label{eqn:J_0_gen}
\end{gather}
\begin{gather}
    {\hat J}_Q  = \frac{4 e t_z }{Q} \sin \frac{Q}{2}\int \frac{d^3 \bm k}{(2\pi)^3} \Big[ \sin\Big(k_z + \frac{Q}{2}\Big) - eA_0(t) \cos\Big(k_z + \frac{Q}{2}\Big) 
    - \frac{1}{2}e^2 A^2_0(t)\sin\Big(k_z + \frac{Q}{2}\Big) \Big] \hat{c}^\dagger_{\bm k }\hat{c}_{\bm k + \bm Q} + \dots\label{eqn:J_Q_gen}
\end{gather}
\end{widetext}
Here the integrals are over the first Brilloin zone (BZ) $k_{\alpha = x,y,z}\in (-\pi,\pi]$, associated with the cubic lattice (we set the lattice constant to one).

\subsection{Inclusion of the CDW order}

If the system is translationally invariant, then a zero momentum electric pump can only generate a zero momentum electric current density. 
The situation becomes more interesting if the translational invariance is spontaneously broken, in which case a zero momentum electric pump can produce currents at $Q = Q_{\rm CDW}$. Here we consider a commensurate CDW order parameter, which we describe as:
\begin{align}
  \hat{H}_{\rm kin} \to  \int' \frac{d^3 \bm k}{(2\pi)^3}  \hat{c}^\dagger_{\bm k, m } (H_0 + H_{\rm CDW})_{mn} \hat{c}_{\bm k,n}, \label{eqn:H_k_CDW}
\end{align}
where the integration now is over the reduced BZ defined by the CDW wave vectors. The commensurability condition is encoded in the fact that $H_{\rm CDW}$ is a finite-dimensional matrix (specific examples are considered in the following sections); for simplicity, we assume this matrix $H_{\rm CDW}$, which encodes the strength of the CDW order parameter, comes from $\hat{H}_{\rm int}$ via mean-field theory. In our notations, the state $\ket{\bm k,n}$ in the reduced BZ corresponds to the state $\ket{\bm k + \bm G_n}$ in the original BZ, where $\bm G_n$ is a wave vector of the reduced reciprocal lattice. For brevity, below we often omit writing the explicit dependence on $\bm k$. The current operators in Eqs.~\eqref{eqn:J_0_gen} and~\eqref{eqn:J_Q_gen} can then be written as:
\begin{align}
    \hat{J}_0  = e \int' \frac{d^3 \bm k}{(2\pi)^3}  \hat{c}^\dagger_{ m } \Big( & J_0^{(0)}  + eA_0(t) J_0^{(1)} \notag\\
    &
    + \frac{1}{2}e^2A^2_0(t) J_0^{(2)} + \dots\Big)_{mn} \hat{c}_{n},\\
    \hat{J}_Q  =  e \int' \frac{d^3 \bm k}{(2\pi)^3}  \hat{c}^\dagger_{ m } \Big( & J_Q^{(0)}  + eA_0(t) J_Q^{(1)} \notag\\
    & 
    + \frac{1}{2}e^2A^2_0(t) J_Q^{(2)} + \dots\Big)_{mn} \hat{c}_{n}.
\end{align}
We now introduce unitary operators $U_{\bm k}$ such that the matrix $U_{\bm k}^\dagger (H_0 + H_{\rm CDW})U_{\bm k}$ becomes diagonal. In the new basis, the operators read:
\begin{widetext}
\begin{gather}
    \hat{H} = \sum_a \int' \frac{d^3 \bm k}{(2\pi)^3} \varepsilon_a  \hat{a}^\dagger_{a } \hat{a}_{a} + \hat{V}_E,\,
    \hat{V}_E =  - eA_0(t) \sum_{ab} \int' \frac{d^3 \bm k}{(2\pi)^3}\hat{a}^\dagger_{a } \Big(h^{(0)} + \frac{1}{2}eA_0(t)h^{(1)} + \dots  \Big)_{ab}  \hat{a}_{b},\label{eqn:Ham_rot}\\
    \hat{J}_0 =  e\sum_{ab} \int' \frac{d^3 \bm k}{(2\pi)^3}\hat{a}^\dagger_{a } \Big(h^{(0)} + eA_0(t)h^{(1)} + \frac{1}{2} e^2A^2_0(t)h^{(2)} + \dots  \Big)_{ab}  \hat{a}_{b},\\
    \hat{J}_Q =  e\sum_{ab} \int' \frac{d^3 \bm k}{(2\pi)^3}\hat{a}^\dagger_{a } \Big( \tilde{h}^{(0)} + eA_0(t)\tilde{h}^{(1)} + \frac{1}{2} e^2A^2_0(t)\tilde{h}^{(2)} + \dots  \Big)_{ab}  \hat{a}_{b},\label{eqn:J_Q_rot}
\end{gather}
\end{widetext}
where $\hat{c}_n = U_{n m}\hat{a}_m$ and $ h^{(0)} \equiv U^\dagger J_0^{(0)} U $, $\tilde{h}^{(0)} = U^\dagger J_Q^{(0)} U$, etc. We remark that in practice, the eigenvalues $\varepsilon_a(\bm k)$, which represent the new Bloch bands in the presence of the CDW order parameter, and unitaries $U_{\bm k}$ are found numerically. Equations~\eqref{eqn:Ham_rot}-\eqref{eqn:J_Q_rot} represent our starting point for evaluating both linear and nonlinear electromagnetic responses perturbatively. 
Below we omit writing the prime sign that indicates the integration is over the reduced BZ.

Let us remark that in developing the presented framework, we are guided by the experiments of Ref.~\cite{stripes1}. Notably, there, photoexcitation of the system with a CDW order, with pump frequency much greater than all energy scales associated with collective modes ($\Omega_{\rm pump}\approx 375\,$THz), results in a low-frequency emission ($\Omega_{\rm out}\simeq 1\,$THz). It implies this downconversion from high-to-low frequencies should be understood from the perspective of mobile electrons rather than slow collective modes, partially justifying our assumption that the former can be described within mean-field theory.

\subsection{Keldysh formulation of the perturbative expansion}
\label{sub:keldysh}

The evaluation of arbitrary response functions of interest can be conveniently carried out using the Keldysh technique~\cite{kamenev2011field}. To this end, we write the partition functional as a path integral in real-time:
\begin{align}
    {\cal Z}[A_0] = \int {\cal D}(\bar{\Psi},{\Psi}) \exp(i {\cal S}[\bar{\Psi},\Psi;A_0] ).
\end{align}
The action is given by:
\begin{align}
    {\cal S} & [\bar{\Psi},\Psi;A_0]   = {\cal S}_0[\bar{\Psi},\Psi] - \text{Tr}\, (\bar{\Psi} V_E\hat{\gamma}^{\rm cl}  \Psi) \\
    &
    = {\cal S}_0[\bar{\Psi},\Psi] -  \int\limits_{-\infty}^{\infty} dt \int \frac{d^3\bm k}{(2\pi)^3} \bar{\Psi}_a(t) (V_E(t))_{ab} \hat{\gamma}^{\rm cl} \Psi_b(t),\notag
\end{align}
where $\Psi_a(\bm k, t)$ and $\bar{\Psi}_a(\bm k, t)$ 
encode the two-component fermionic fields that live in the real-time Keldysh space.
In the second line, we indicated that the trace involves the integration over both time and momentum and summation over the Bloch bands. 
Here ${\cal S}_0[\bar{\Psi},\Psi]$ is the unperturbed quadratic action, fully characterized by the bare Green's function $\hat{G}_{\alpha\beta} (t,t') \equiv -i \langle \Psi_\alpha(t) \bar{\Psi}_\beta(t')\rangle_0$, which can be conveniently written in the Keldysh space as
\begin{align}
    \hat{G}_{\alpha\beta} (t,t') = \begin{pmatrix}
       G^R(t,t') & G^K(t,t')\\
       0 & G^A(t,t')
    \end{pmatrix}. 
\end{align}
Here $G^R$, $G^A$, and $G^K$ are the usual retarded, advanced, and Keldysh Green's functions, respectively. 
In the basis of Eq.~\eqref{eqn:Ham_rot}, they are given by:
\begin{gather}
    G^{R(A)}_a(\omega) = \frac{1}{\omega - \varepsilon_a \pm i\eta},\\
    G^K_a(\omega) = -2\pi iF(\omega)\delta(\omega - \varepsilon_a),
\end{gather}
where $F(\omega) = 1 - 2n_F(\omega)$, $n_F(\omega)$ is the Fermi-Dirac function. Below, we are interested in computing observables of the type ${\cal O}(t) =  \langle {\cal O}_{ab} \hat{a}^\dagger_a(t)\hat{a}_b(t)\rangle$, which in the Keldysh technique are understood as:
\begin{align*}
{\cal O}(t) = \int {\cal D}(\bar{\Psi},{\Psi}) \exp(i {\cal S}[\bar{\Psi},\Psi;A_0] ) \,
 \frac{1}{2}{\cal O}_{ab} \bar{\Psi}_a(t) \hat{\gamma}^{\rm q}\Psi_b(t).
\end{align*}
In the equations above, $\hat{\gamma}^{\rm cl} = \hat{I}$ and $\hat{\gamma}^{\rm q} = \sigma_x$ are the standard $2\times 2$ matrices in the Keldysh space.

We proceed perturbatively in the amplitude of the external pump field $A_0(t)$ and primarily focus on evaluating the Umklapp current $J_Q(t)$:
\begin{widetext}
\begin{align}
    J_Q(t) = \frac{e}{2}\Big\langle  \Bar{\Psi}(t)\Big(\tilde{h}^{(0)} + eA_0(t)\tilde{h}^{(1)} & + \frac{1}{2} e^2A^2_0(t)\tilde{h}^{(2)} + \dots  \Big)\hat{\gamma}^{\rm q}{\Psi}(t)\notag\\
    & \times \Big( 1 - i \text{Tr}\, (\bar{\Psi} V_E\hat{\gamma}^{\rm cl}  \Psi) - \frac{1}{2}[\text{Tr}\, (\bar{\Psi} V_E\hat{\gamma}^{\rm cl}  \Psi)]^2 + \dots \Big) \Big\rangle_0,\label{eqn:gen_J_Q_pert}
\end{align}
\end{widetext}
where the average here is with respect to the bare action ${\cal S}_0[\bar{\Psi},\Psi]$. Since the latter is only quadratic in the fermionic fields $\Psi$ and $\bar{\Psi}$, the expectation values of any observables can be computed using Wick's theorem.

\section{Linear response}
\label{sec:linear}

The above framework formally allows us to compute the linear-response Umklapp current, which might be interesting on its own. To the leading order in $A_0(t)$, we get two contributions from Eq.~\eqref{eqn:gen_J_Q_pert}:
\begin{align}
    J^{(1)}_Q(t) & = \frac{e^2 A_0(t)}{2}\Big\langle \Bar{\Psi}(t)\tilde{h}^{(1)} \hat{\gamma}^{\rm q}{\Psi}(t) \Big\rangle_0 \notag\\
    & 
    \qquad\qquad
    - \frac{i e}{2} \Big\langle \Bar{\Psi}(t)\tilde{h}^{(0)} \hat{\gamma}^{\rm q}{\Psi}(t) \text{Tr}\, (\bar{\Psi} V^{(1)}_E\hat{\gamma}^{\rm cl}  \Psi) \Big) \Big\rangle_0 \notag\\
    & = e^2 A_0(t) \sum_a \int \frac{d^3\bm k}{(2\pi)^3} f_a \tilde{h}^{(1)}_{aa} \notag\\
    & \qquad\qquad\qquad\qquad
    + e^2 \int d t' \, \Pi(t,t') A_0(t'),
\end{align}
where 
\begin{align}
    \Pi(t,t')  = & \frac{i}{2}\sum_{ab} \int \frac{d^3\bm k}{(2\pi)^3} h^{(0)}_{ab} \tilde{h}^{(0)}_{ba}  
    \\
    &
    \times (G^R_a(t,t')G^K_b(t',t) + G^K_a(t,t')G^A_b(t',t) ). \notag 
\end{align}
Using $\bm E_0(t) = -\partial_t \bm A_0(t)$, we obtain the linear conductivity:
\begin{align}
     \sigma^{(1)}_Q(\omega) & =  \frac{e^2}{i\omega}\Big(  \sum_a \int \frac{d^3\bm k}{(2\pi)^3} f_a \tilde{h}^{(1)}_{aa} + \Pi(\omega) \Big)\notag\\
    &  =\frac{e^2}{i\omega}\Big(  \sum_a \int \frac{d^3\bm k}{(2\pi)^3} f_a \tilde{h}^{(1)}_{aa} \notag\\
    &\qquad\qquad\quad
    + \sum_{ab} \int \frac{d^3\bm k}{(2\pi)^3} \frac{ h^{(0)}_{ab} \tilde{h}^{(0)}_{ba} f_{ab}}{\omega - \varepsilon_{ab} + i\eta} \Big). \label{eqn:sigma_LR}
\end{align}
Here $f_{ab}= f_a - f_b$ is the difference in the fermionic occupation numbers; $\varepsilon_{ab} = \varepsilon_{a} -\varepsilon_{b}$ is the energy difference between the corresponding Bloch bands.
Equation~\eqref{eqn:sigma_LR} is the central result of this section, and it states that in systems with a CDW order parameter, zero momentum electric field can generate a nonzero Umklapp current; within linear response, the frequency of this current is the same as that of the external perturbation. Strictly speaking, following the philosophy outlined in the previous section, the result in Eq.~\eqref{eqn:sigma_LR} is valid only at large frequencies; at low frequencies, one will have to take into account collective charged excitations and go beyond the mean-field approach. We remark that the regular conductivity $\sigma^{(1)}_0(\omega)$, associated with the zero momentum electric current, is obtained from Eq.~\eqref{eqn:sigma_LR} by replacing $\tilde{h}$ with $h$. In the next section, we turn to compute the second-order Umklapp current.

\section{Second-order response and The Umklapp shift current}
\label{sec:nonlinear}

\subsection{General results}

To the second order in $A_0(t)$, Eq.~\eqref{eqn:gen_J_Q_pert} gives four terms:
\begin{widetext}
\begin{align}
    J_Q^{(2)}(t)  = &  \frac{e^3 A^2_0(t)}{4}\Big\langle \Bar{\Psi}(t)\tilde{h}^{(2)} \hat{\gamma}^{\rm q}{\Psi}(t) \Big\rangle_0  - \frac{i e^2A_0(t)}{2} \Big\langle \Bar{\Psi}(t)\tilde{h}^{(1)} \hat{\gamma}^{\rm q}{\Psi}(t) \text{Tr}\, (\bar{\Psi} V^{(1)}_E\hat{\gamma}^{\rm cl}  \Psi) \Big) \Big\rangle_0\notag\\
    &
    - \frac{ie}{2}\Big\langle  \Bar{\Psi}(t)\tilde{h}^{(0)}\hat{\gamma}^{\rm q}{\Psi}(t)  \text{Tr}\, (\bar{\Psi} V^{(2)}_E\hat{\gamma}^{\rm cl}  \Psi)\Big\rangle_0 - \frac{e}{4}\Big\langle  \Bar{\Psi}(t)\tilde{h}^{(0)}\hat{\gamma}^{\rm q}{\Psi}(t) [\text{Tr}\, (\bar{\Psi} V^{(1)}_E\hat{\gamma}^{\rm cl}  \Psi)]^2   \Big\rangle_0. \label{eqn:J_2_v1}
\end{align}
Proceeding similarly as above and using Wick's theorem, we obtain the expression for the second-order Umklapp conductivity:
\begin{align}
    \sigma_Q^{(2)}(\omega;\omega_1,\omega_2) = & \frac{e^3}{i\omega_1 i\omega_2} \sum_{abc}\int \frac{d^3\bm k}{(2\pi)^3} \Big[  f_a \tilde{h}^{(2)}_{aa}  +  \frac{ h^{(0)}_{ab} \tilde{h}^{(1)}_{ba} f_{ab}}{\omega_2 - \varepsilon_{ab} + i\eta} +   \frac{ h^{(0)}_{ab} \tilde{h}^{(1)}_{ba} f_{ab}}{\omega_1 - \varepsilon_{ab} + i\eta}
    +   \frac{ h^{(1)}_{ab} \tilde{h}^{(0)}_{ba} f_{ab}}{\omega - \varepsilon_{ab} + i\eta}\notag
    \\
    &  + 
    h^{(0)}_{ab}h^{(0)}_{bc} \tilde{h}^{(0)}_{ca} \frac{1}{\omega - \epsilon_{ac} + i\eta} 
    \Big(
      \frac{f_{ab}}{\omega_1 - \epsilon_{ab} + i\eta} + \frac{f_{ab}}{\omega_2 - \epsilon_{ab} + i\eta} - \frac{f_{bc}}{\omega_1 - \epsilon_{bc} + i\eta} - \frac{f_{bc}}{\omega_2 - \epsilon_{bc} + i\eta}
    \Big)
    \Big].\label{eqn:full_sigma_Q_2}
\end{align}
Equation~\eqref{eqn:full_sigma_Q_2} is the most generic answer for the second-order electromagnetic response, from which, in particular, we get the conductivity associated with the Umklapp shift current:
\begin{align}
     \sigma_Q^{(2)}(0;\omega,-\omega) = &
    \frac{e^3}{\omega^2} \sum_{ab, c\neq a} \int \frac{d^3\bm k}{(2\pi)^3} \Big[  f_a \tilde{h}^{(2)}_{aa}  +  \frac{ h^{(0)}_{ab} \tilde{h}^{(1)}_{ba} f_{ab}}{\omega - \varepsilon_{ab} + i\eta} +   \frac{ h^{(0)}_{ab} \tilde{h}^{(1)}_{ba} f_{ab}}{-\omega - \varepsilon_{ab} + i\eta}
    -   \frac{ h^{(1)}_{ab} \tilde{h}^{(0)}_{ba} f_{ab}}{ \varepsilon_{ab}} \notag\\
     & - 
    h^{(0)}_{ab}h^{(0)}_{bc} \tilde{h}^{(0)}_{ca} \frac{1}{\epsilon_{ac}} 
    \Big(
      \frac{f_{ab}}{\omega - \epsilon_{ab} + i\eta} + \frac{f_{ab}}{-\omega - \epsilon_{ab} + i\eta} - \frac{f_{bc}}{\omega - \epsilon_{bc} + i\eta} - \frac{f_{bc}}{-\omega - \epsilon_{bc} + i\eta}
    \Big)
    \Big].\label{eqn:sigma_shiftU}
\end{align}
\end{widetext}
Equation~\eqref{eqn:sigma_shiftU} is the main result of this section, which states that a high-frequency zero momentum pump electric field can be downconverted to a low-frequency Umklapp shift current. We remark that in obtaining Eq.~\eqref{eqn:sigma_shiftU}, i.e., when taking the limit of the outgoing frequency to zero in Eq.~\eqref{eqn:full_sigma_Q_2}, it is understood that the limit $\eta \to 0$ is taken first. Phenomenologically, $\eta$ partially accounts for the effects of the short-range quenched disorder in the sample. When the disorder is notable and the spectral width of the incident radiation is narrow, one instead might want to consider the limit of the outgoing frequency to zero first. In this case, one starts from Eq.~\eqref{eqn:full_sigma_Q_2} and carefully carries out the corresponding evaluation.

The expression for the regular shift current~\cite{Parker2019Diagrammatic} is obtained from Eq.~\eqref{eqn:sigma_shiftU} via replacing all matrices $\tilde{h}$ with $h$. In the literature~\cite{Aversa1995,Sipe2000,cook2017design}, a more standard expression is well-known, which was derived using the length gauge:
\begin{align}
    \sigma^{(2)}_{\alpha \beta \beta}(0;\omega,-\omega)   = & 2 \pi e^3 \int \frac{d^3 {\bm k}}{(2\pi)^3} \sum_{ab} f_{ab}\\
    & \times \text{Im}\Big(r^\beta_{ba}r^\beta_{ab;\alpha} \Big) \delta(\varepsilon_{ab} - \omega), \label{eqn:sigma2_gen_lg}\notag
\end{align}
where indices $\alpha$ and $\beta$ represent the Cartesian directions; $r_{ab}^\alpha (\bm k)= v^\alpha_{ab}(\bm k)/i\varepsilon_{ab}(\bm k)$ for $a \neq b$ and zero otherwise; $v^\alpha_{ab}(\bm k) = \Bra{a}\partial_{k_\alpha} H(\bm k)\Ket{b}$ encodes the velocity operator matrix elements. We also introduced~\cite{Aversa1995,Sipe2000,cook2017design}:
\begin{align*}
     r^\alpha_{ab;\beta}  = & -\frac{r^\alpha_{ab} \Delta^\beta_{ab} + r^\beta_{ab} \Delta^\alpha_{ab}}{\varepsilon_{ab}} + \frac{w^{\alpha\beta}_{ab}}{i\varepsilon_{ab}} \\
     & - \frac{1}{\varepsilon_{ab}} \sum_{c\neq a,b} (v^\alpha_{ac}r^\beta_{cb} - r^\beta_{ac} v^\alpha_{cb}) ,\text{ for } a\neq b,
\end{align*}
where $\Delta^\alpha_{ab} = v_{aa}^\alpha-v_{bb}^\alpha$ and $w^{\alpha\beta}_{ab}(\bm k) = \Bra{a}\partial_{k_\alpha}\partial_{k_\beta} H(\bm k)\Ket{b}$. Of course, the two expressions for the regular zero momentum shift currents in the two gauges agree with each other (we numerically checked this statement), but to see this explicitly requires a nontrivial application of sum rules that relate the two gauges to each other (see, for example, Appendix~B of Ref.~\cite{chaudhary2021shift} for details).

\begin{figure}[b!]
\centering
\includegraphics[width=1\linewidth]{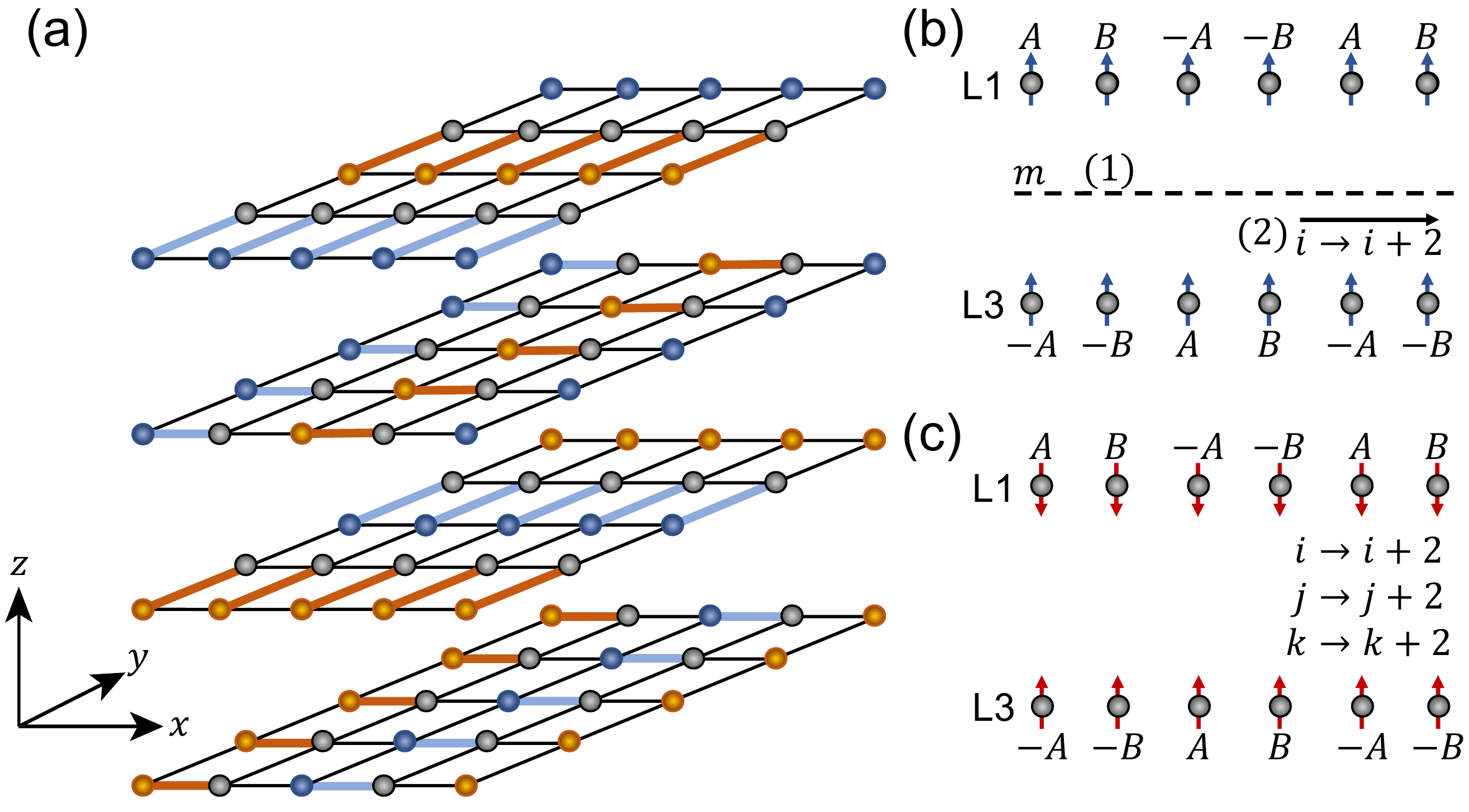} 
\caption{(a) Schematic of period four helical-like stripes in layered materials~\cite{RevModPhys.75.1201}. A bond-centered pattern is described by alternating weak (blue), unmodulated (solid black), and strong (orange) bonds. Likewise, a site-centered pattern is characterized by alternating sites with depleted (blue), unmodulated (grey), and accumulated (orange) electron density. (b) Illustration of a symmetry that hinders the development of the regular $z$-axis shift current $J_0$. The shown symmetry corresponds to the reflection with respect to layer 2 (dashed line) with subsequent translation by two lattice sites along the $x$-axis. Here we consider a site-centered pattern, where different letters encode local charge density modulation. Blue arrows indicate the direction of $J_0$. (c) An example of a symmetry, consisting of three lattice translations along $x$-, $y$-, and $z$-axes, that hinders the development of the Umklapp shift current $J_U$ (its direction is shown with red arrows). }
\label{fig::Period_4_stripes}
\end{figure}

\subsection{Relation to experiments}
\label{subsec:exp}

We turn to discuss how the Umklapp shift currents can be observed in experiments. The momentum of such currents is set by the period of the CDW order, while their frequency is close to zero. This puts them far outside the light cone, and, as such, one expects radiation generated by these currents to be evanescent outside the sample. The decay length of the electric field in the air should be comparable to the wavelength. Thus generically, detection of an Umklapp current requires local probes, such as SNOM detectors \cite{RevModPhys.84.1343}.

Remarkably, in certain systems, it can also be detected with far-field experiments, similarly to how the regular shift current has been observed in non-centrosymmetric materials~\cite{Xu.2019a,Moore.2013,Tan.2016,Auston.1972,Glass.1974,Somma.2014}. A possible recent example of such a situation is the terahertz emission experiments in striped superconductors~\cite{stripes1}. Specifically, under the assumption that strong photoexcitation gives rise to nonzero Umklapp shift current, it was theoretically predicted~\cite{stripes2} that this current acts as a drive to long-wavelength surface Josephson plasmons. These excitations, in turn, can emit light, and the far-field experimental
observation of this emission was suggested as evidence for the Umklapp shift current~\cite{stripes1,stripes2}.

\section{Analysis and discussion of striped layered materials}
\label{sec:striped}

\subsection{Stripes patterns with the lowest momentum harmonics}

Here we apply the results of the previous section and investigate the regular and Umklapp shift currents in layered materials with stripes. For concreteness, we consider the following microscopic model:
\begin{align}
    \hat{H} = \hat{H}_0 + \hat{H}_{\rm CDW},
\end{align}
where $\hat{H}_0$ is the effective kinetic energy of the system without stripes. We choose the simplest possible dispersion, which in momentum space reads:
\begin{align}
    E_0(\bm k) = - 2t (\cos k_x + \cos k_y)- 2t_z \cos k_z - \mu.
\end{align}
One instead could choose this dispersion to better mimic cuprates, such as the one in Ref.~\cite{Pavarini2001}; in this work, we aim to understand the electrodynamics of striped systems from the generic perspective of lattice symmetries rather than to match a specific model of high-temperature superconductors. Let us remark that while below we investigate stripe patterns, our framework can be equally applied to analyze checkerboards~\cite{RevModPhys.75.1201,wise2008charge,comin2015symmetry,PhysRevLett.98.197003,PhysRevB.75.014205,PhysRevB.75.060504,PhysRevB.74.134507,Dolgirev2017}. We write the CDW part of the Hamiltonian as $\hat{H}_{\rm CDW} = \hat{H}_b + \hat{H}_s$, where $\hat{H}_b$ and $\hat{H}_s$ encode bond-centered and site-centered modulations [see Fig.~\ref{fig::Period_4_stripes}(a)], respectively. We concentrate on commensurate CDWs, relevant for hole doping close to $1/8$, in which case the stripes have period four in each direction~\cite{Tranquada.1996,RevModPhys.75.1201,laliberte2011fermi,PhysRevB.84.012507,comin2015broken,jang2016ideal,PhysRevLett.99.127003}. In particular, the most generic Hamiltonian for stripes corresponding to the lowest momentum harmonics, with $Q_x = Q_y = Q_z = \pi/2$, takes the following form:
\begin{align}
    \hat{H}_b & = V_b \sum_{ijk} \Big\{ \sin \Big( \frac{\pi}{2}k \Big) \sin\Big(\frac{\pi}{2} i + \vartheta_{b}^x\Big) \hat{c}^\dagger_{i,j,k}\hat{c}_{i + 1,j,k}  
    \notag\\
    &
    +  \cos \Big( \frac{\pi}{2}k \Big) \sin\Big(\frac{\pi}{2} j + \vartheta_{b}^y\Big)\hat{c}^\dagger_{i,j,k}\hat{c}_{i,j+1,k} + h.c.\Big\} ,\label{eqn:H_b}
\end{align}
\begin{align}
    \hat{H}_s & = V_s \sum_{ijk} \Big\{ \sin \Big( \frac{\pi}{2}k \Big) \sin\Big(\frac{\pi}{2} i + \vartheta_{s}^x\Big) \notag\\
    &
    + \cos \Big( \frac{\pi}{2}k \Big) \sin\Big(\frac{\pi}{2} j + \vartheta_{s}^y\Big)\Big\} \hat{c}^\dagger_{i,j,k}\hat{c}_{i,j,k}.\label{eqn:H_s}
\end{align}
As illustrated in Fig.~\ref{fig::Period_4_stripes}(a), a system with such a pattern forms a chiral structure, which in turn breaks inversion symmetry, except for special cases such as $\vartheta_b^x  =\pi m$
($\vartheta_s^x =  \pi m$) 
or $\vartheta_b^y =  \frac{\pi}{2} + \pi n$ 
($\vartheta_s^y = \frac{\pi}{2} + \pi n$), where 
$m,n\in \mathds{Z}$.

\begin{figure*}[hbt!]
\centering
\includegraphics[width=1\linewidth]{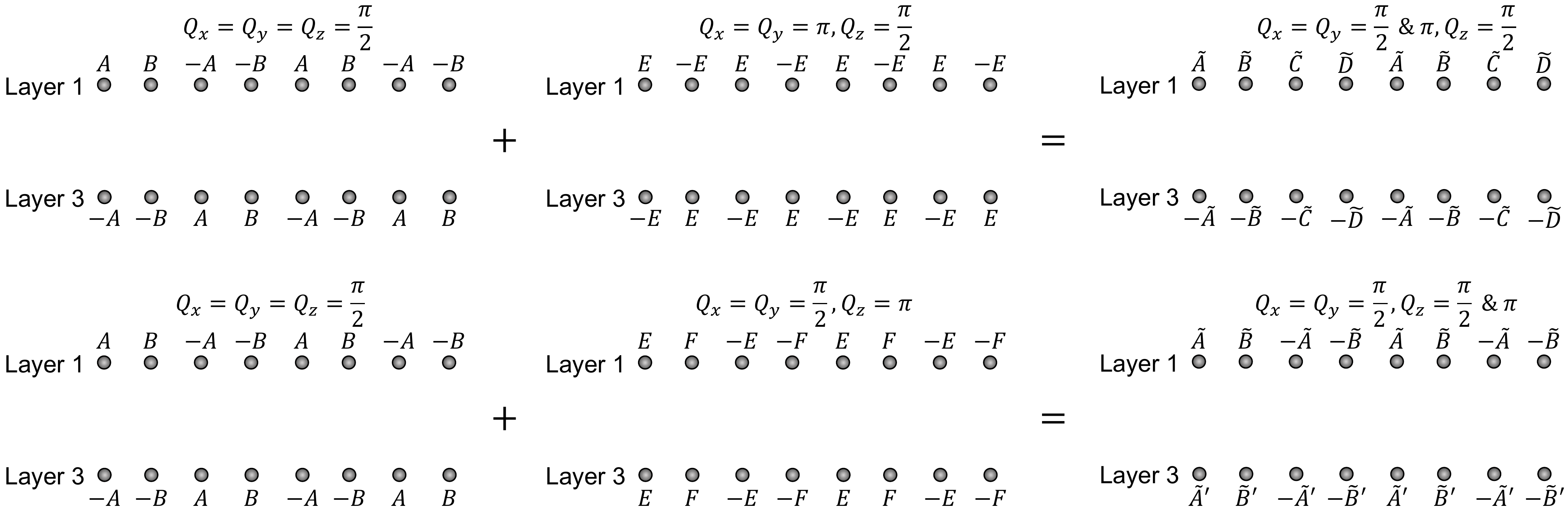} 
\caption{ Illustration that a generic superposition of the lowest momenta harmonics with higher ones can break all the hindering lattice symmetries that do not allow for the development of a nonzero shift current $J_{{\rm shift},z}$. }
\label{fig::Higher_harmonics}
\end{figure*}
{\renewcommand{\arraystretch}{1.7}

The question we want to address is whether the system with such a modulation can host regular and/or Umklapp shift currents:
\begin{align*}
    J_{{\rm shift}, z} = J_0 + J_U(z) = J_0 + (J_{Q} e^{iQz} + c.c.),\, Q = \frac{\pi}{2}.
\end{align*}
The answer to this question is negative: while the stripes patterns in Eqs.~\eqref{eqn:H_b} and~\eqref{eqn:H_s} can break inversion symmetry, they still have remnant symmetries that enforce $J_0 = J_U = 0$. An example of such a symmetry for the regular shift current is shown in Fig~\ref{fig::Period_4_stripes}(b): Under the action of $S = {T}^x_2 \circ {R}_{z}$, both $\hat{H}_b$ and $\hat{H}_s$ or their arbitrary superposition remain unchanged, but $J_0\to -J_0$ changes sign, which is compatible with ${S}$ being a symmetry only if $J_0 = 0$. We defined  ${R}_{z}$ as the reflection with respect to the $xy$-plane (so that $k \to -k$) and ${T}^x_2$ as the translation operator by two sites along the $x$-axis ($i \to i+2$). An example of a symmetry, ${T}^z_2\circ{T}^y_2\circ{T}^x_2$, that hinders the Umklapp shift current $J_U$ is illustrated in Fig.~\ref{fig::Period_4_stripes}(c). We remark that to be a little bit more precise, one should also take into account the presence of the external zero momentum electric field. For the linear response conductivities, this electric field can be thought of as the source that breaks inversion symmetry and, as such, allows for nonzero electric currents. In contrast, the shift currents are second-order processes and, as such, do not depend on whether the external electric field points upwards or downwards so that the above lattice symmetry analysis becomes highly relevant.

\subsection{The role of higher momentum harmonics}

Does this mean that stripes alone cannot give rise to the shift current $J_{{\rm shift}, z}$? While the patterns with the lowest momentum harmonics cannot give nonzero $J_{{\rm shift}, z}$, we find that the inclusion of higher momentum harmonics will be sufficient to break all the hindering symmetries, as we turn to discuss. We remark that these higher momentum harmonics are expected to be generically present; however, diffraction and STM experiments in cuprates~\cite{Tranquada.1996,abbamonte2005spatially,da2014ubiquitous,comin2015symmetry,comin2015broken,jang2016ideal} indicate that their amplitudes are expected to be weak.

We consider three classes of stripes with higher harmonics: i) those with period two along the $z$-axis ($Q_z = \pi$) and period four within the layers ($Q_x = Q_y= \pi/2$); ii) those with period four along the $z$-axis ($Q_z = \pi/2$) and period two in each of the layers ($Q_x = Q_y= \pi$); iii) those with period two, both in-plane and out-of-plane ($Q_x = Q_y= Q_z = \pi$). These patterns can be represented as:
 \begin{align}
     \hat{H}_{b}^{z} & = {V}^z_b \sum_{ijk} \Big\{ \frac{1 - \cos (\pi k) }{2} \sin\Big(\frac{\pi}{2} i + \alpha_b^x\Big) \hat{c}^\dagger_{i,j,k}\hat{c}_{i + 1,j,k}\notag\\
    & + 
    \frac{1 + \cos (\pi k) }{2} \sin\Big(\frac{\pi}{2} j + \alpha_b^y\Big)\hat{c}^\dagger_{i,j,k}\hat{c}_{i,j + 1,k} + h.c. \Big\} \label{eqn:Hb_z} ,\\
     \hat{H}_{s}^{z}  & =  {V}^z_s\sum_{ijk} \Big\{ \frac{1 - \cos (\pi k) }{2} \sin\Big(\frac{\pi}{2} i + \alpha_s^x\Big) \notag\\
    &
    +
    \frac{1 + \cos (\pi k) }{2} \sin\Big(\frac{\pi}{2} j + \alpha_s^y\Big)\Big\} \hat{c}^\dagger_{i,j,k}\hat{c}_{i,j,k},\label{eqn:Hs_z}
\end{align}
\begin{align}
     \hat{H}_{b}^{xy}  & = {V}^{xy}_b \sum_{ijk} \Big\{ \sin \Big( \frac{\pi}{2}k \Big) \sin(\pi i + \beta_b^x) \hat{c}^\dagger_{i,j,k}\hat{c}_{i + 1,j,k} \notag\\ 
     &
     + 
   \cos \Big( \frac{\pi}{2}k \Big) \sin(\pi j + \beta_b^y\Big)\hat{c}^\dagger_{i,j,k}\hat{c}_{i,j + 1,k} + h.c. \Big\} ,\label{eqn:Hb_ab}\\
    \hat{H}_{s}^{xy} & =  {V}^{xy}_s\sum_{ijk} \Big\{ \sin \Big( \frac{\pi}{2}k \Big) \sin(\pi i + \beta_s^x)  \notag\\
    &
    + \cos \Big( \frac{\pi}{2}k \Big) \sin(\pi j + \beta_s^y)\Big\} \hat{c}^\dagger_{i,j,k}\hat{c}_{i,j,k},\label{eqn:Hs_ab}
\end{align}
\begin{align}
    \hat{H}_b^{(2)} & = \tilde{V}_b \sum_{ijk} \Big\{ \frac{1 - \cos (\pi k) }{2} \sin(\pi i + \varphi_b^x) \hat{c}^\dagger_{i,j,k}\hat{c}_{i + 1,j,k} \notag\\
    & 
    + 
    \frac{1 + \cos (\pi k) }{2} \sin(\pi j + \varphi_b^y)\hat{c}^\dagger_{i,j,k}\hat{c}_{i,j + 1,k} + h.c. \Big\},\label{eqn:Hb2}\\ 
     \hat{H}_{s}^{(2)} & =  \tilde{V}_s \sum_{ijk} \Big\{ \frac{1 - \cos (\pi k) }{2} \sin(\pi i + \varphi_s^x) \notag\\
     &
     +
    \frac{1 + \cos (\pi k) }{2} \sin(\pi j + \varphi_s^y)\Big\} \hat{c}^\dagger_{i,j,k}\hat{c}_{i,j,k}.\label{eqn:Hs2}
\end{align}
In Appendix~\ref{appendix_Mat_elems}, we compute the corresponding matrix elements used in our numerical analyses below.

\begin{figure}[t!]
\centering
\includegraphics[width=1\linewidth]{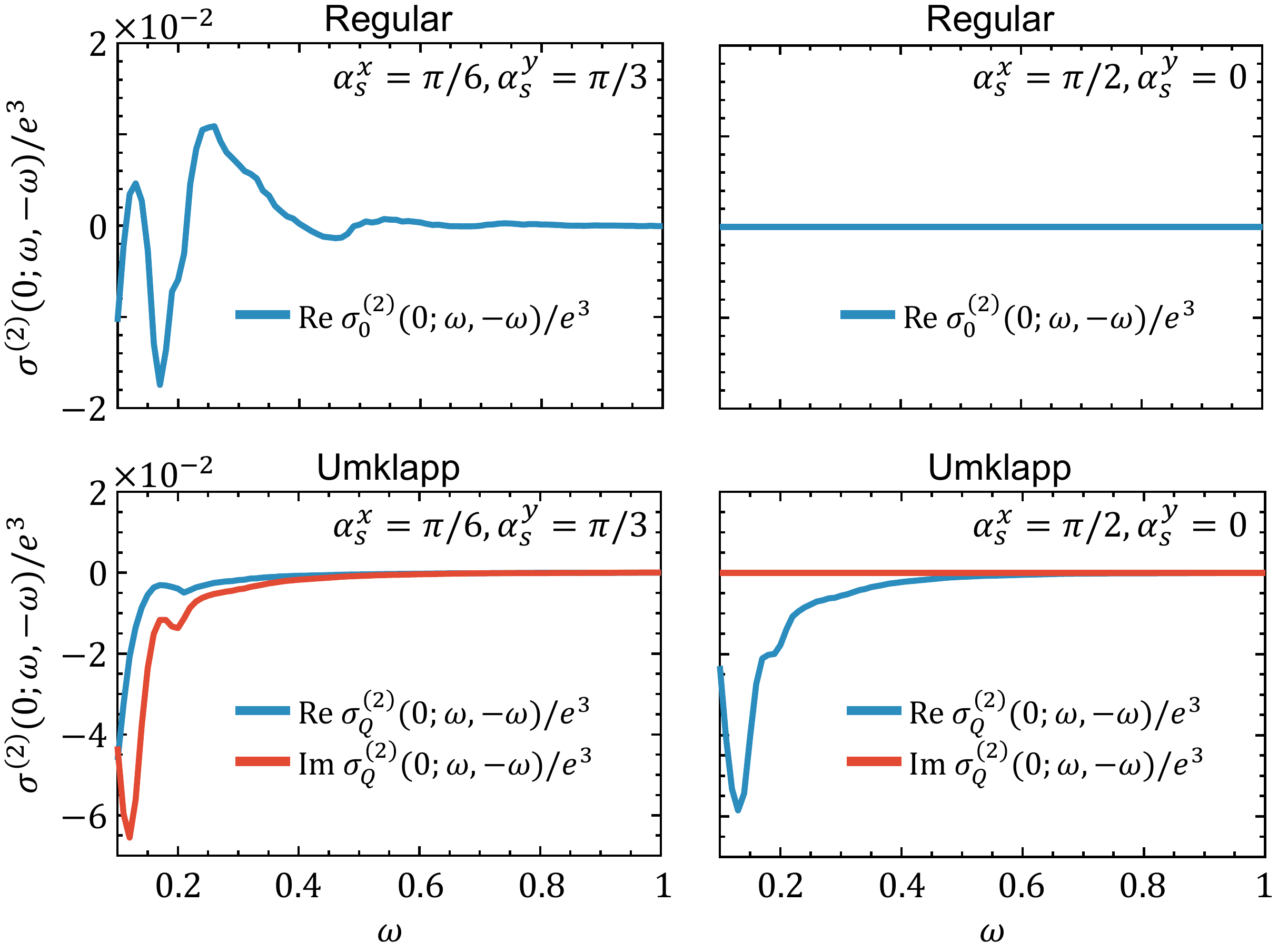} 
\caption{ Microscopic evaluation of the regular (top panels) and Umklapp (bottom panels) shift currents in layered materials with stripes. We find that only in the presence of higher momentum harmonics, those currents are in general nonzero (left panels). In case there is a screw axis, it enforces the regular shift current to be zero (top right panel); at the same time, the Umklapp component can be nonzero (bottom right panel). Parameters used: $t_z = 0.25$ (here we fix $t = 1$), $V_s = 0.4$, $V_s^z = 0.2$, $T = 0.1$; if a parameter is not specified, it means it is set to zero; the chemical potential $\mu$ is such that it mimics the hole doping to be $1/8$; we choose sufficiently large system size $512\times512\times512$ to ensure a decent convergence of the shown results. }
\label{fig::Currents_stripes}
\end{figure}

Explicit case by case investigation shows that likewise the charge patterns with the lowest momentum harmonics, either of these new classes alone cannot give rise to nonzero shift currents. An example of a symmetry that hinders $J_0$ in the first class is ${R}_{z}$ and that hinders $J_U$ is ${T}_2^z$. The other two classes can be analyzed similarly. However, as illustrated in Fig.~\ref{fig::Higher_harmonics}, a generic superposition of the lowest momenta patterns with either of the first two classes does break all the remnant symmetries resulting in both $J_0\neq 0$ and $J_U \neq 0$. Indeed, an example of a microscopic calculation, cf. Eq.~\eqref{eqn:sigma_shiftU}, confirming this conclusion, is shown in Fig.~\ref{fig::Currents_stripes} (left panels). Let us also mention that a generic admixture of the lowest momenta stripes with the third class is not sufficient to give rise to either of the shift currents because it turns out that both ${S}$ and ${T}^z_2\circ{T}^y_2\circ{T}^x_2$ are also symmetries of Eqs.~\eqref{eqn:Hb2} and~\eqref{eqn:Hs2}. Finally, we note that a generic admixture of stripes patterns from different classes, although not very relevant for cuprates, can give rise to both $J_0\neq 0$ and $J_U\neq 0$. These conclusions, which we explicitly checked numerically, are summarized in Table~\ref{tab::summary}.

\begin{table}[t!]
\large
\centering
\begin{tabularx}{\linewidth}{| Y | Y | Y | Y | Y | Y | Y | }
\hline
\hline
$Q_x$ & $\frac{\pi}{2}$ & $\frac{\pi}{2}$ & $\pi$ & $\pi$ & $\frac{\pi}{2}$ & $\frac{\pi}{2} + \pi$ \\
$Q_y$ & $\frac{\pi}{2}$ & $\frac{\pi}{2}$ & $\pi$ & $\pi$ & $\frac{\pi}{2}$ & $\frac{\pi}{2} + \pi$ \\
$Q_z$ & $\frac{\pi}{2}$ & $\pi$ & $\frac{\pi}{2}$ & $\pi$ & $\frac{\pi}{2} + \pi$ & $\frac{\pi}{2}$ \\
\hline
$J_{0}$ & \xmark  & \xmark & \xmark & \xmark & \text{\color{cadmiumgreen}{ \cmark}$^\dagger$} &  \text{\color{cadmiumgreen}{ \cmark}$^\dagger$} \\
$J_{U}$ & \xmark  & \xmark & \xmark & \xmark & \text{\color{cadmiumgreen}{ \cmark}} &  \text{\color{cadmiumgreen}{ \cmark}} \\
\hline
\hline
\end{tabularx}
\caption{Summary of the analysis of lattice symmetries of various stripes patterns. Provided there is an admixture of the lowest momenta harmonics with higher momenta ones, one can get nonzero $z$-axis shift currents $J_0\neq 0$ and $J_U \neq 0$. $\dagger$ requires in addition that all screw axes are broken.}
\label{tab::summary}
\end{table}
}

\begin{figure}[t!]
\centering
\includegraphics[width=1\linewidth]{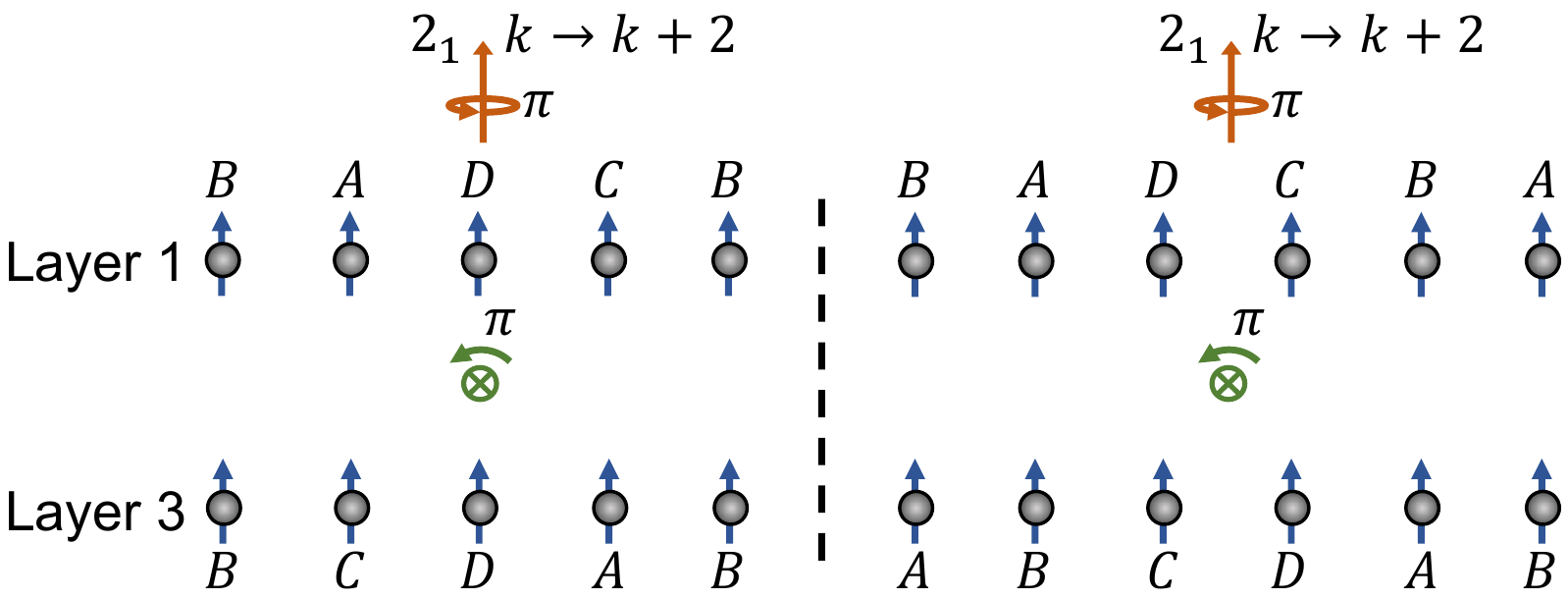} 
\caption{ Two examples where the CDW pattern has a $2_1$ screw axis (shown in orange). In this case, the system turns out to be symmetric with respect to a rotation by $\pi$ around an axis parallel to the $y$-axis (shown in green). Under such a rotation, the regular shift current transforms as $J_0 \to -J_0$ (its direction is encoded in blue arrows), which is compatible with this rotation being a symmetry only if $J_0 = 0$.}
\label{fig::Screw_axis}
\end{figure}

So far, we have considered a generic situation where the phases that appear in Eqs.~\eqref{eqn:H_b}-\eqref{eqn:Hs2} are not tuned to some specific values. Implicit in our construction of the stripes patterns is that the CDWs in adjacent layers are orthogonal to each other, and the corresponding momentum harmonics have the same amplitudes. Despite this, unless the phases of those harmonics are appropriately adjusted, the system is not C$_4$-like symmetric. Seeking simple patterns with a helical structure, one might want to impose a $4_1$-screw axis symmetry so that the stripes in adjacent layers are obtained from each other by a $\pi/2$ rotation around the $z$-axis. Such a symmetry, in turn, implies $2_1$-screw axis so that the stripes in the next-to-nearest layers are related to each other through rotation by $\pi$. Notably, as we demonstrate in Fig.~\ref{fig::Screw_axis}, this latter symmetry is incompatible with the regular shift current $J_0$, enforcing $J_0 = 0$. In contrast, this symmetry does not constrain the Umklapp shift current $J_U$ so that, in principle, it can be nonzero $J_U \neq 0$, as we show in Fig.~\ref{fig::Currents_stripes} (right panels). In Appendix~\ref{appendix_symmetries}, we summarize all point-group symmetries that allow for the regular shift current (the point-group symmetry of stripes in Fig.~\ref{fig::Screw_axis} is ``422'', which gives $J_0 = 0$, although inversion symmetry can be broken).

\subsection{Inversion symmetry and the Umklapp shift current}
\label{sub:symmetry}

\begin{figure}[t!]
\centering
\includegraphics[width=1\linewidth]{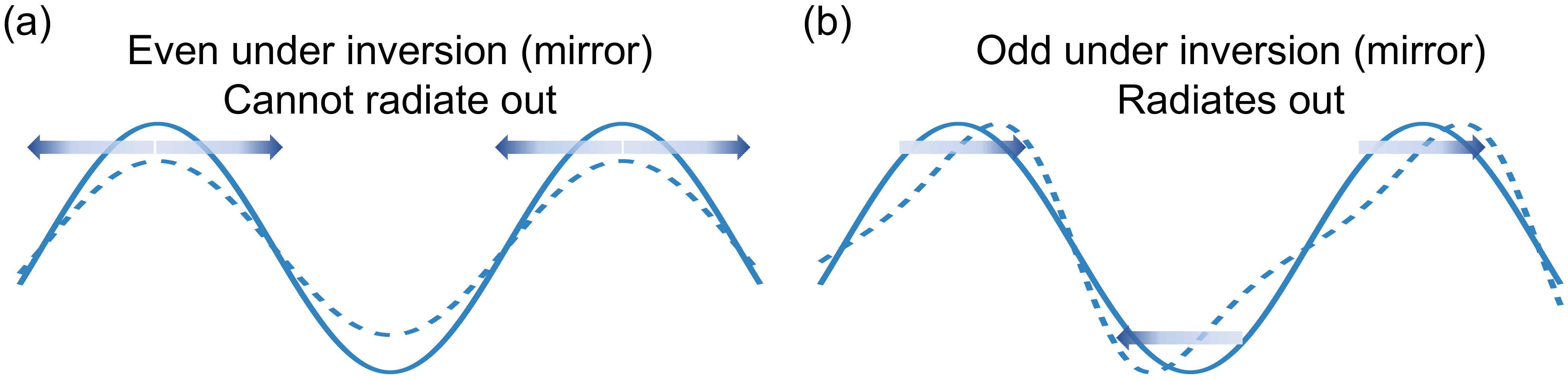} 
\caption{Inversion symmetry and Umklapp currents. (a) Cartoon of a situation corresponding to a distortion of the CDW amplitude, which, in general, will lead to nonzero Umklapp currents (shown with blue arrows). In this case, the resulting pattern of currents is even under inversion, which, in turn, implies  that the system will not emit light. (b) Phase distortions of the CDW order also result in Umklapp currents, which can be odd under inversion; shown pattern of currents, therefore, will emit light.}
\label{fig::Inversion}
\end{figure}

This observation that the regular shift current is zero in helical stripes, but the Umklapp shift current is nonzero, is surprising. Another important such example is in systems with inversion symmetry. Since inversion symmetry breaking is a prerequisite for the regular shift current, we get $J_0 = 0$. At the same time, the Umklapp current is, in general, nonzero $J_U \neq 0$ -- see Fig.~\ref{fig::Inversion}(a) for an illustration, where the CDW amplitude is partially evaporated so that the system remains inversion symmetric but develops nonzero Umklapp currents. The challenge one might need to resolve experimentally is how to detect such an Umklapp current, given that the pattern of currents in Fig.~\ref{fig::Inversion}(a), due to being even under inversion, cannot radiate light out. In contrast, the current pattern in Fig.~\ref{fig::Inversion}(b), since it is odd under inversion, can emit light and, thus, is detectable with far-field probes~\cite{stripes1,stripes2}.

\subsection{Implications for cuprates}
\label{sub:cuprates}

While the full complexity of high-temperature cuprate superconductors is well beyond the presented modeling, our symmetry analysis still provides valuable insights about the origin of the regular and/or Umklapp shift currents in these materials. Most importantly, we found that the stripe patterns with the lowest momenta harmonics cannot give rise to either of these currents. At the same time, diffraction experiments indicate~\cite{Tranquada.1996} that higher momentum stripes, illustrated in Fig.~\ref{fig::Higher_harmonics}, are suppressed. Therefore, one possibility is that the experiments of Ref.~\cite{stripes1} are much more sensitive to these higher harmonics. Another aspect that should be considered is that in some of the compounds investigated in the experiments~\cite{stripes1}, the CDW patterns are actually incommensurate and, as such, all the hindering symmetries are expected to be broken (especially given that incommensurate stripes are susceptible to quenched disorder). Finally, an interesting alternative is that the superconducting and charge orders intertwine to form a putative PDW order predicted in Ref.~\cite{Berg_2009}. In this case, it is argued that due to the frustrated $\pi$-Josephson couplings inherent in the PDW state, a form of non-collinear phase ordering emerges, which then also breaks all of the hindering symmetries and gives rise to both shift currents.

We turn to briefly remark on the bulk origin of the shift currents in striped superconductors~\cite{stripes1,stripes2}. We note that despite lattice symmetries (such as inversion symmetry) hindering bulk optical rectification in the simplest models of stripes, one could argue that these symmetries are broken at the surface of the material, which potentially could provide an alternative explanation of the experiment. In this scenario, however, optically stimulated terahertz radiation would be observed in LBCO samples regardless of the presence of the stripe order. This is different from the experimental results presented in Ref.~\cite{stripes1}, which indicate only a weak broadband radiation above the charge transition temperature. The strong outgoing radiation, with the frequency being sharply peaked at the Josephson plasma resonance, was seen only below superconducting $T_c$ in samples with stripes. Therefore, bulk rectification, which we study here, is an essential ingredient in understanding the experiment.

\section{Conclusion and Outlook}
\label{sec:conclusion}

We comment briefly on the finite momentum nature of the response we have considered. To date, studies of nonlinear optics have largely focused on the zero momentum response -- for two good reasons. First, the momentum transferred by an optical photon is typically negligible compared to the Brillouin zone size and, second, any currents that do happen to arise at large momenta are generally outside the light cone, and thus are expected to decay before they may be observed. We have shown that a CDW order may evade these restrictions. Bragg scattering of quasiparticles off the charge order naturally gives rise to currents at CDW reciprocal lattice momenta and, under certain lattice symmetries, the Umklapp shift current is the dominant low-frequency effect. Furthermore, there are mechanisms, such as the surface Josephson plasmon physics discussed above, that enable detection of the Umklapp shift current in the far-field regime~\cite{stripes2}.

For the outlook, there are several promising directions. On the theoretical front, it would be interesting to: i) establish whether the Umklapp shift current has a topological origin, as has been recently discussed in the context of the regular nonlinear probes~\cite{morimoto2016topological}; ii) extend the presented framework to incorporate the effects of the static short-range disorder, as many strongly-correlated materials have substantial randomness; iii) include the effects of interactions beyond mean-field~\cite{Lee.2021} to understand the role of low-energy collective excitations better~\cite{dolgirev2021periodic}
; and iv) generalize our modeling to include such phases as PDW states. On the experimental side, additional probes, such as Raman scattering~\cite{SaiHalasz.1978,Klein.1982,Kupcic.2009,GarciaRuiz.2020} and near-field measurements~\cite{RevModPhys.84.1343}, that might sense both shift currents are highly desirable, as they can help confirm our theoretical conclusions.

\section*{ACKNOWLEDGEMENTS}
The authors would like to thank M.~D. Lukin, M. Mitrano, P.~I. Arseev, A. Zong, I. Esterlis, and A. Radkevich for fruitful discussions. P.E.D., M.H.M., and E.D. were supported by AFOSR-MURI: Photonic Quantum Matter award FA95501610323, DARPA DRINQS, and the ARO grant``Control of Many-Body States Using Strong Coherent Light-Matter Coupling in Terahertz Cavities''. This research is funded in part by the Gordon and Betty Moore Foundation’s EPiQS Initiative, Grant GBMF8683 to D.E.P. J.B.C. is supported by the Quantum Science Center (QSC), a National Quantum Information Science Research Center of the U.S. Department of Energy (DOE), the ARO grant ``Control of Many-Body States Using Strong Coherent Light-Matter Coupling in Terahertz Cavities'', and by the Harvard Quantum Initiative. J.B.C. also acknowledges hospitality from the Max Planck Institute for Structure and Dynamics of Matter (MPSD, Hamburg), and ETH Zürich Institute for Theoretical Physics. D.N., M.B., M.F., and A.C. were supported by the European Research Council under the European Union’s Seventh Framework Programme (FP7/2007-2013)/ERC Grant Agreement No. 319286 (QMAC), and the Deutsche Forschungsgemeinschaft (DFG, German Research Foundation) via the excellence cluster ‘The Hamburg Centre for Ultrafast Imaging’ (EXC 1074 – project ID 194651731), and the priority program SFB925 (project ID 170620586).

\bibliography{Sc}

\onecolumngrid
\appendix

\section{Matrix elements of various CDW patterns}
\label{appendix_Mat_elems}

Here we summarize the matrix elements of the stripes patterns discussed in the main text:
\begin{align}
    \Bra{a^\prime,b^\prime,c^\prime} & \hat{H}_b  \Ket{a,b,c}  = -\frac{ V_b}{4} [e^{i\vartheta_b^x}\delta_{a-a^\prime + 1} - e^{-i\vartheta_b^x}\delta_{a-a^\prime - 1}] [e^{i(k_x + \frac{\pi}{2}a)} + e^{-i(k_x + \frac{\pi}{2}a^\prime)}] \times \delta_{b-b^\prime} \times  [\delta_{c-c^\prime + 1} - \delta_{c-c^\prime - 1}] \notag\\
    & - \frac{i V_b}{4} \delta_{a-a^\prime}  \times [e^{i\vartheta_b^y}\delta_{b-b^\prime + 1} - e^{-i\vartheta_b^y}\delta_{b-b^\prime - 1}][e^{i(k_y + \frac{\pi}{2}b)} + e^{-i(k_y + \frac{\pi}{2}b^\prime)}] \times  [\delta_{c-c^\prime + 1} + \delta_{c-c^\prime - 1}],\\
    \Bra{a^\prime,b^\prime,c^\prime} & \hat{H}_s  \Ket{a,b,c}  =  -\frac{ V_s}{4} [e^{i\vartheta_s^x}\delta_{a-a^\prime + 1} - e^{-i\vartheta_s^x}\delta_{a-a^\prime - 1}] \times \delta_{b-b^\prime} \times  [\delta_{c-c^\prime + 1} - \delta_{c-c^\prime - 1}] \notag\\
    & - \frac{iV_s}{4} \delta_{a-a^\prime}  \times [e^{i\vartheta_s^y}\delta_{b-b^\prime + 1} - e^{-i\vartheta_s^y}\delta_{b-b^\prime - 1}] \times  [\delta_{c-c^\prime + 1} + \delta_{c-c^\prime - 1}],\\
    \Bra{a^\prime,b^\prime,c^\prime} & \hat{H}_{b}^{z}  \Ket{a,b,c}  =  -\frac{ i V^z_b}{8} [e^{i\alpha_b^x}\delta_{a-a^\prime + 1} - e^{-i\alpha_b^x}\delta_{a-a^\prime - 1}] [e^{i(k_x + \frac{\pi}{2}a)} + e^{-i(k_x + \frac{\pi}{2}a^\prime)}] \times \delta_{b-b^\prime} \times [2 \delta_{c - c^\prime} - \delta_{c-c^\prime + 2} - \delta_{c-c^\prime - 2}] \notag\\
    & -\frac{ i V^z_b}{8} \delta_{a-a^\prime}  \times [e^{i\alpha_b^y}\delta_{b-b^\prime + 1} - e^{-i\alpha_b^y}\delta_{b-b^\prime - 1}] [e^{i(k_y + \frac{\pi}{2}b)} + e^{-i(k_y + \frac{\pi}{2}b^\prime)}] \times   [2 \delta_{c - c^\prime} + \delta_{c-c^\prime + 2} + \delta_{c-c^\prime - 2}],\\
    \Bra{a^\prime,b^\prime,c^\prime} & \hat{H}_{s}^{z}  \Ket{a,b,c}  =  -\frac{ i V^z_s}{8} [e^{i\alpha_s^x}\delta_{a-a^\prime + 1} - e^{-i\alpha_s^x}\delta_{a-a^\prime - 1}] \times \delta_{b-b^\prime} \times [2 \delta_{c - c^\prime} - \delta_{c-c^\prime + 2} - \delta_{c-c^\prime - 2}] \notag\\
    & -\frac{ i V^z_s}{8} \delta_{a-a^\prime}  \times [e^{i\alpha_s^y}\delta_{b-b^\prime + 1} - e^{-i\alpha_s^y}\delta_{b-b^\prime - 1}] \times   [2 \delta_{c - c^\prime} + \delta_{c-c^\prime + 2} + \delta_{c-c^\prime - 2}],\\
    \Bra{a^\prime,b^\prime,c^\prime} &  \hat{H}_{b}^{xy}  \Ket{a,b,c}  =  -\frac{ V_b^{xy}}{4} [e^{i\beta_b^x}\delta_{a-a^\prime + 2} - e^{-i\beta_b^x}\delta_{a-a^\prime - 2}] [e^{i(k_x + \frac{\pi}{2}a)} + e^{-i(k_x + \frac{\pi}{2}a^\prime)}] \times \delta_{b-b^\prime} \times  [\delta_{c-c^\prime + 1} - \delta_{c-c^\prime - 1}] \notag\\
    & - \frac{iV_b^{xy}}{4} \delta_{a-a^\prime}  \times [e^{i\beta_b^y}\delta_{b-b^\prime + 2} - e^{-i\beta_b^y}\delta_{b-b^\prime - 2}][e^{i(k_y + \frac{\pi}{2}b)} + e^{-i(k_y + \frac{\pi}{2}b^\prime)}] \times  [\delta_{c-c^\prime + 1} + \delta_{c-c^\prime - 1}],\\
    \Bra{a^\prime,b^\prime,c^\prime} & \hat{H}_{s}^{xy}  \Ket{a,b,c}  =  -\frac{  V_s^{xy}}{4} [e^{i\beta_s^x}\delta_{a-a^\prime + 2} - e^{-i\beta_s^x}\delta_{a-a^\prime - 2}] \times \delta_{b-b^\prime} \times  [\delta_{c-c^\prime + 1} - \delta_{c-c^\prime - 1}] \notag\\
    & - \frac{i V_s^{xy}}{4} \delta_{a-a^\prime}  \times [e^{i\beta_s^y}\delta_{b-b^\prime + 2} - e^{-i\beta_s^y}\delta_{b-b^\prime - 2}] \times  [\delta_{c-c^\prime + 1} + \delta_{c-c^\prime - 1}],\\
    \Bra{a^\prime,b^\prime,c^\prime} & \hat{H}_{b}^{(2)}  \Ket{a,b,c} = -\frac{ i \tilde{V}_b}{8} [e^{i\varphi_b^x}\delta_{a-a^\prime + 2} - e^{-i\varphi_b^x}\delta_{a-a^\prime - 2}] [e^{i(k_x + \frac{\pi}{2}a)} + e^{-i(k_x + \frac{\pi}{2}a^\prime)}] \times \delta_{b-b^\prime} \times [2 \delta_{c - c^\prime} - \delta_{c-c^\prime + 2} - \delta_{c-c^\prime - 2}] \notag\\
    & -\frac{ i \tilde{V}_b}{8} \delta_{a-a^\prime}  \times [e^{i\varphi_b^y}\delta_{b-b^\prime + 2} - e^{-i\varphi_b^y}\delta_{b-b^\prime - 2}] [e^{i(k_y + \frac{\pi}{2}b)} + e^{-i(k_y + \frac{\pi}{2}b^\prime)}] \times   [2 \delta_{c - c^\prime} + \delta_{c-c^\prime + 2} + \delta_{c-c^\prime - 2}],\\
    \Bra{a^\prime,b^\prime,c^\prime} & \hat{H}_{s}^{(2)}  \Ket{a,b,c}  =  -\frac{ i \tilde{V}_s}{8} [e^{i\varphi_s^x}\delta_{a-a^\prime + 2} - e^{-i\varphi_s^x}\delta_{a-a^\prime - 2}] \times \delta_{b-b^\prime} \times [2 \delta_{c - c^\prime} - \delta_{c-c^\prime + 2} - \delta_{c-c^\prime - 2}] \notag\\
    & -\frac{ i \tilde{V}_s}{8} \delta_{a-a^\prime}  \times [e^{i\varphi_s^y}\delta_{b-b^\prime + 2} - e^{-i\varphi_s^y}\delta_{b-b^\prime - 2}] \times   [2 \delta_{c - c^\prime} + \delta_{c-c^\prime + 2} + \delta_{c-c^\prime - 2}],
\end{align}
where the equalities in the Kronecker symbols are up to modulo 4. Here $a,b,c\in \{0,1,2,3\}$; the state $\Ket{a,b,c}$ implicitly encodes momentum $\bm k$ in the reduced BZ, with $k_x,k_y,k_z \in (-\pi/4,\pi/4]$, and momentum $\bm k + \frac{\pi}{2}a \hat{e}_x + \frac{\pi}{2}b \hat{e}_y + \frac{\pi}{2}c \hat{e}_z$ in the original BZ. For a fixed momentum $\bm k$ in the reduced BZ, the size of any matrix is $64\times 64$.

\section{Symmetry analysis for optical rectification}
\label{appendix_symmetries}

We analyze point groups that allow for optical rectification utilizing the tools of the Bilbao Crystallographic Database~\cite{bilbaosymmetry2019} -- the results are summarized in Tab.~\ref{tab::symmetry}. In addition, there we mark point groups that can give rise to the regular zero momentum shift current when both the incoming field and outgoing polarization are parallel to the $z$-axis.

\begin{table}
\centering
\begin{tabular}{| M{4.5cm}| M{4.5cm}| M{4.5cm} | }
\hline
Point group &  Number of independent elements of $\chi(0;\omega,-\omega)$  & $\chi_{33}$ \\ \hline
1 & 10 & \text{\color{cadmiumgreen}{ \cmark}} \\
2 & 4 & \xmark \\ 
m & 6 & \text{\color{cadmiumgreen}{ \cmark}} \\
222 & 1 & \xmark \\ 
mm2 & 3 & \text{\color{cadmiumgreen}{ \cmark}}\\ 
4 & 2 & \text{\color{cadmiumgreen}{ \cmark}} \\ 
-4 & 3 & \xmark \\ 
4mm & 2 & \text{\color{cadmiumgreen}{ \cmark}} \\ 
-42m & 1 & \xmark \\ 
3 & 6 & \text{\color{cadmiumgreen}{ \cmark}} \\ 
32 & 2 & \xmark \\ 
3m & 4 & \text{\color{cadmiumgreen}{ \cmark}} \\ 
6 & 2 & \text{\color{cadmiumgreen}{ \cmark}} \\ 
-6 & 4 & \xmark \\ 
6mm & 2 & \text{\color{cadmiumgreen}{ \cmark}} \\ 
-6m2 & 2 & \xmark \\ 
23 & 1 & \xmark \\ 
-43m & 1 & \xmark \\ \hline
\end{tabular}
\caption{ Left: list of point groups that allow for optical rectification. Center: number of independent elements in $\chi(0;\omega,-\omega)$ for each of these groups. Right: point groups marked in green can give rise to the regular shift current along the
$z$-axis when the incident light is also parallel to the $z$-axis.}
\label{tab::symmetry}
\end{table}

\end{document}